\documentclass[prd,preprint,superscriptaddress,showpacs,byrevtex]{revtex4} 
\usepackage{epsfig} 
 
\def\as{\alpha_s} 
\def\gl{\tilde{g}} 
\def\sq{\tilde{q}} 
\def\sqb{\bar{\tilde{q}}} 
\def\qb{\bar{q}}

\def\ms{m_{\tilde q}} 
\def\mg{m_{\tilde g}}

\def\md{m_{-}}

\def\ghat{\hat{g}_s}

\def\bs{\beta_{\sq}}

\def\bg{\beta_{\gl}}

\newcommand{\Lg}[2]{\log\left(\frac{#1}{#2}\right)}

\newcommand{\be}{\begin{equation}} 
\newcommand{\ee}{\end{equation}} 
\newcommand{\bea}{\begin{eqnarray}} 
\newcommand{\eea}{\end{eqnarray}} 
\def\simgt{\rlap{\lower 3.5 pt \hbox{$\mathchar \sim$}} \raise 1pt 
Ê \hbox {$>$}} 
 
\def\as{\alpha_s} 
 
\def\ms{m_{\tilde q}} 
\def\mg{m_{\tilde g}} 
\def\md{m_{-}} 

\begin{document} 

\title{SUSY Production From TeV Scale Blackhole at LHC} 

\author{Andrew Chamblin} \email{chamblin@lns.mit.edu} 
\affiliation{Department of Physics, University of Louisville, Louisville, KY 40292, USA} 

\author{Fred Cooper} \email{fcooper@nsf.gov} 
\affiliation{ National Science Foundation, Arlington, VA 22230, USA and T-8, Theoretical Division, 
Los Alamos National Laboratory, Los Alamos, NM 87545, USA} 

\author{Gouranga C. Nayak} \email{nayak@insti.physics.sunysb.edu} 
\affiliation{ C. N. Yang Institute for Theoretical Physics, Stony Brook University, SUNY, Stony Brook, 
NY 11794-3840, USA } 

\date{\today} 

\begin{abstract} 
If the fundamental Planck scale is near a TeV, then we should
expect to see TeV scale black holes at the LHC.
Similarly, if the scale of supersymmetry breaking is sufficiently 
low, then we might expect to see light supersymmetric particles in the next 
generation of colliders. If the mass of the supersymmetric particle is of order  
a TeV and is comparable to the temperature of a typical TeV scale 
black hole, then such sparticles will be copiously 
produced via Hawking radiation: 
The black hole will act as a resonance for 
sparticles, among other things. In this paper we compared various signatures for SUSY 
production at LHC, and we contrasted the situation where the sparticles are produced 
directly via parton fusion processes
with the situation where they are produced indirectly through black hole resonances. 
We found that black hole resonances provide a larger source for heavy mass SUSY 
(squark and gluino) production than the direct pQCD-SUSY production via parton 
fusion processes depending on the values of the Planck mass and blackhole mass.
Hence black hole production at LHC may indirectly act as a dominant channel for SUSY production. 
We also found that the differential cross section $d\sigma/dp_t$ for 
SUSY production increases as a function of the $p_t$ (up to $p_t$ equal to about 1
 TeV or more) of the SUSY particles (squarks and
gluinos), which is in sharp contrast with the pQCD predictions where the differential cross section
$d\sigma/dp_t$ decreases as $p_t$ increases for high $p_t$ about 1 TeV
or higher. This is a feature for any 
particle emission from TeV scale blackhole as long as the
temperature of the blackhole is very high ($\sim $ TeV). 
Hence measurement of increase of $d\sigma/dp_t$ with 
$p_t$ for $p_t$ up to about 1 TeV or higher
for final state particles might be a useful signature for blackhole production at LHC.

\end{abstract} 
\pacs{PACS: } %
\maketitle 

\newpage 

\section{Introduction} 

It is now generally accepted that the scale of quantum gravity {\it could be} as low as a TeV \cite{folks}.
Also, it is unknown if supersymmetry is manifested in nature, and if SUSY does exist it is unclear at what energy scale SUSY becomes manifest. 
Pragmatically,  we can only search for black holes, extra dimensions and superpartners in whatever mass range is currently accessible to 
experiment.  If we manage to detect {\it any} of these exotic phenomena, then we will be propelled into the twenty-first 
century, as our understanding of quantum gravity and perhaps even string theory is revolutionized.
There are many discussions of graviton, radion and black hole production at 
LHC. If such processes occur they will probe 
TeV scale quantum gravity at the collider experiments \cite{gravr,gravr1}. 
One of the most exciting aspects of this TeV scale gravity will be the production of black holes in particle accelerators. These 
`brane-world' black holes will be our first window into the extra dimensions of space predicted by string theory, and required by the several brane-world 
scenarios that provide for a low energy Planck scale \cite{large}. 
Using various approximations, in a number of recent papers people have studied the production of microscopic black holes in proton-proton 
(pp) and lead-lead (PbPb) collisions at LHC and cosmic ray events
\cite{ppbf,pp,pp1,pp2,pp3,ag,ppch,ppk,ppu,park,hof,more,cham,cooper,pp4,pp5,pp6}. 
Typically, it only makes sense to say that a `black hole' has formed at 
several times the Planck scale - anything smaller will dissolve into 
something known as string ball \cite{sball}.
Thus, if the lightest supersymmetric particle (LSP) is sufficiently lighter 
than the black hole temperature, then we expect that such sparticles will be 
produced as the black hole evaporates through the Hawking process \cite{lands}. Naively, 
the lighter the LSP, 
and the larger the Planck scale then the more sparticles we expect to see produced through this process. However, as the Planck scale is increased the production 
cross section of the blackhole in pp collisions at LHC is decreased. Hence, SUSY production comes from two competitive effects as the Planck scale is 
increased: 1) the SUSY production from a single blackhole increases 2) the 
cross section for a single blackhole to be produced decreases. Similarly, if the SUSY scale is right near the Planck scale, then the rate 
for sparticle production through this channel is minimal. In this paper we perform a systematic analysis which contrasts SUSY (squark and gluino) production from 
direct pQCD-SUSY production processes with SUSY emission from a 
blackhole via Hawkins radiation 
at LHC. We find that squark and gluino production from 
a blackhole at LHC can be larger or smaller than direct squark and 
gluino production from pQCD-SUSY processes 
depending on the value of the TeV scale Planck mass, 
the number of extra dimensions and the black hole mass. 
We find that as long as the temperature of the blackhole is of the order of a TeV, the squark and gluino production cross sections from the 
blackhole do not depend too much on the squark and gluino masses. On the otherhand the direct pQCD-SUSY production cross sections 
seriously depend on the masses of the squark and gluino, since in 
any massive particle production using pQCD calculations  the cross section depends on the mass explicitly. 
This provides us with an important conclusion:  If TeV scale blackholes are indeed formed at LHC, then one signature of this will be an unusually 
copious production of massive particles, which can not be formed through direct fusion processes. Hence if we observe very high rates of massive particle 
production at LHC, it might provide indirect evidence that TeV scale blackholes are being produced at LHC. We make a detailed analysis 
of this in this paper in the context of massive squark and gluino production. 
We also study the differential cross section for squark and gluino
production both from the blackhole and from the pQCD-SUSY processes.
One of the interesting results we find is that as long as the
temperature of the blackhole is very high ($\sim$ 1 TeV) then 
the differential cross section $d\sigma/dp_t$ increases
as $p_t$ is increased (up to about $p_t$ equal to 1 TeV or more). 
This is in sharp contrast to pQCD predictions
where $d\sigma/dp_t$ decreases as $p_t$ is increased when one is at  high $p_t$. This is
not only true for SUSY particles but also true for any particles
emitted from  a blackhole via Hawking radiation as long as the temperature
of the black hole is high ($\sim$ TeV). This is
very interesting because if one experimentally observes at LHC that
the $d\sigma/dp_t$ for final state particles
increases as $p_t$ is increased (up to about 1 TeV or higher)
then it might provide a good signature for blackhole production at LHC. We make a 
detailed analysis of all these observations in this paper.

The paper is organized as follows: 
In section II we present the calculation of the 
differential and total cross sections for squark and gluino production 
at LHC in a typical MSSM scenario, 
where we take the gluino to be the LSP. In section III we present a 
computation for the rate of squark and gluino production from a blackhole
via Hawking radiation. In section IV we present the results and a discussion. 

\section{Squark and Gluino Production in pp Collisions at LHC using pQCD } 

Here we briefly discuss the SUSY (squark and gluino) production mechanism
in quark and gluon fusion processes using pQCD methods applied to high
energy hadronic collisions. The production channels for squarks ($\tilde{q}, \tilde{\bar q}$) 
and gluinos ($\tilde{g}$) in pp collisions at the leading order (LO) of the perturbative 
expansion are given by: $q \bar q, ~gg \rightarrow {\tilde q} {\tilde{\bar q}}, 
~{\tilde{g} \tilde{g}}$, $q \bar q \rightarrow {\tilde q} {\tilde{q}}$, $q g 
\rightarrow {\tilde q} {\tilde{g}}$ etc.  The Feynman diagrams for these 
processes are shown in Fig. 1.
For the production of squark pairs or squark--gluino pairs only one initial state contributes at 
lowest order. For squark--antisquark and gluino pairs both gluon--gluon and quark--antiquark 
initial states are included. The 
differential cross section in the lowest order is given by:
\bea
\frac{d\sigma}{dp_t^2 dy}~=~\frac{H}{s}~ &&
\sum_{i,j=q,\bar q, g} ~\int_{x_1^{min}}^1 dx_1 ~(- \frac{1}{x_1^2t})
~f_{i/A}(x_1,p_t^2)~f_{j/B}(-\frac{x_1 t s}{x_1s+u},p_t^2)  \times \nonumber \\
&&~\sum 
|M|^2(\frac{-x_1^2ts}{x_1s+u},x_1t,-\frac{x_1tu}{x_1s+u})
\eea
where $s$ is the total center of mass energy at hadronic level,
$t=-\sqrt{s(p_t^2+m^2)}~e^y$ and $u=-\sqrt{s(p_t^2+m^2)}~e^{-y}$. The
lower limit of the $x_1$ integration is given by $x_1^{min}=-u/s+t$. 
$f_{i/p}(x,Q^2)$ is the parton distribution function inside a proton with 
longitudinal momentum fraction $x$ and factorization scale $Q$. 
$H=K_{ij}/16\pi$ with $K_{qq}=K_{q\bar q}=1/36$, $K_{gg}=1/256$ and
$K_{qg}=1/96$. The matrix element squares from partonic fusion processes
at the Born level (see Fig-1) are given by \cite{been}: 

\begin{small}
\begin{eqnarray*}
  \sum |{\cal M}^B|^2 (q_i\qb_j\to\sq\sqb) & = & 
  \delta_{ij}\left[32 n_f g_s^4  \frac{\hat t_q \hat u_q -\ms^2 \hat s}{\hat s^2} 
  + 16 \ghat^4  \frac{\hat t_q \hat u_q -(\ms^2 -\mg^2) \hat s}{\hat t_g^2}
\right. \label{born1} \\ 
  & &\hphantom{\delta_{ij}a} \left. {} \hspace*{-1cm}
  - 32/3 g_s^2 \ghat^2 \frac{\hat t_q \hat u_q -\ms^2 \hat s}{\hat s \hat t_g}
\right] 
  + (1-\delta_{ij}) 
  \left[ 16 \ghat^4 \frac{\hat t_q \hat u_q -(\ms^2 -\mg^2)\hat s}{\hat t_g^2}\right]
    \nonumber\\[0.2cm] 
    \sum |{\cal M}^B|^2 (gg\to\sq\sqb) & = & \!\!4 n_f g_s^4
    \left[24\!\left(1 -2\,\frac{\hat t_q \hat u_q}{\hat s^2}\right) -\! 8/3\right]\! \left[ 1
      -\!2\,\frac{\hat s\ms^2}{\hat t_q \hat u_q}\left(1 -\!\frac{\hat s\ms^2}{\hat t_q
        \hat u_q}\right)\right] \quad\qquad\\[0.2cm] 
  \label{squarkborn}
  \sum |{\cal M}^B|^2 (q_i q_j\to\sq\sq) & = & 
  \delta_{ij}\Bigg[ 8\ghat^4 \left(\hat t_q \hat u_q -\ms^2 \hat s\right)
  \left(\frac{1}{\hat t_g^2} +\frac{1}{\hat u_g^2}\right)  \\ 
  & &{} \hphantom{\delta_{ij}a} + 16\ghat^4 \,\mg^2  \hat s \left(  \left(\frac{1}{\hat t_g^2} +\frac{1}{\hat u_g^2}\right)
  - 8/3  \,\frac{1}{\hat t_g \hat u_g}\right) \Bigg]\nonumber\\
  & &{} + (1-\delta_{ij}) 
  \left[ 16 \ghat^4 \ \,\frac{\hat t_q \hat u_q -(\ms^2-\mg^2) \hat s}{\hat t_g^2}\right]
    \nonumber 
\end{eqnarray*}
\begin{eqnarray*}
  \sum |{\cal M}^B|^2 (q\qb\to\gl\gl) & = & 
  96 g_s^4 \, \left[\frac{2 \mg^2 \hat s +\hat t_g^2 +\hat u_g^2}{\hat s^2} \right]\\
  & &{} + 96 g_s^2 \ghat^2 \, \left[\frac{\mg^2 \hat s +\hat t_g^2}{\hat s \hat t_q}
  +\frac{\mg^2 \hat s +\hat u_g^2}{\hat s \hat u_q} \right ]\nonumber\\ 
  & &{} + 2\ghat^4 \left[ 24 \left(\frac{\hat t_g^2}{\hat t_q^2} +
  \frac{\hat u_g^2}{\hat u_q^2}\right) +8/3 \left(2\,\frac{\mg^2 \hat s}{\hat t_q\,\hat u_q} -
  \frac{\hat t_g^2}{\hat t_q^2} - \frac{\hat u_g^2}{\hat u_q^2} \right)\right]\nonumber\\[0.2cm] 
  \sum |{\cal M}^B|^2 (gg\to\gl\gl) & = & 576g_s^4   \left(1
  -\frac{\hat t_g \hat u_g}{\hat s^2}\right) \left[\frac{\hat s^2}{\hat t_g\,\hat u_g} -2 
  +4\,\frac{\mg^2 \hat s}{\hat t_g \hat u_g} \left (1-\frac{\mg^2 \hat s}{\hat t_g \hat u_g}\right)
  \right]\\[0.2cm]    
  \sum |{\cal M}^B|^2 (qg\to\sq\gl) & = & 
  2  g_s^2 \ghat^2 \left[ 24 \left( 1 -2\,\frac{\hat s \hat u_q}{\hat t_g^2}
  \right) - 8/3 \right] 
  \Bigg[-\frac{\hat t_g}{\hat s} \label{born6} \\
  & &{} + \frac{2(\mg^2-\ms^2)\,\hat t_g}{\hat s \hat u_q} \left(1 +\frac{\ms^2}{\hat u_q}
  +\frac{\mg^2}{\hat t_g} \right) \Bigg]. \nonumber
\end{eqnarray*}
\end{small}
In the above equation the variables $\hat t_{q,g}$ ($\hat u_{q,g}$) are related to
the Mandelstam variables $\hat t$ ($\hat u$) by: 
$\hat t_q~(\hat u_q)$~=~$\hat t~(\hat u)~-~m_{\tilde q}^2$
and
$\hat t_g~(\hat u_g)$~=~$\hat t~(\hat u)~-~m_{\tilde g}^2$ where
$m_{\tilde q}$ ($m_{\tilde g}$) is the mass of the squark (gluino). $g_s$ is the QCD
gauge coupling ($qqg$) and $\hat g_s$ is the Yukawa coupling ($q {\tilde q}{\tilde g}$).
$n_f$ is the number of flavors. In the above equations:  
\be
\hat s ~=~-\frac{x_1^2ts}{x_1s+u}~~~~~~~\hat t_{g(q)}~=~x_1t~~~~~~~~\hat u_{g(q)}~=~-\frac{x_1tu}{x_1s+u}.
\ee

The $p_t$ distributions for squarks and gluinos at LHC are determined
from the above formula by integrating over the rapidity:
\be
\frac{d\sigma}{dp_t}~=~2 p_t~\int_{-cosh^{-1}(\frac{\sqrt s}{2(\sqrt{p_t^2
+m^2}})}^{cosh^{-1}(\frac{\sqrt s}{2(\sqrt{p_t^2+m^2}})}
~dy~\frac{d\sigma}{dm_t^2dy}.
\ee
Similarly, the rapidity distributions for squarks and gluinos produced at LHC are 
given by by integrating over $p_t$:
\be
\frac{d\sigma}{dy}~=~ \int_{m^2}^{s/4cosh^2y}~dm_t^2~\frac{d\sigma}{dm_t^2dy}.
\ee

The total cross sections for squarks and gluinos produced at LHC are then given by:
\begin{equation} 
\sigma^{pp \rightarrow {\tilde q} {\tilde {\bar q}} ({\tilde g}{\tilde g})} = 
\Sigma_{i,j} \int_{4M^2/s} dx_1 \int_{4m^2/sx_1} dx_2 f_{i/p}(x_1,Q^2) f_{j/p}(x_2,Q^2) 
\sigma^{ij}(\hat s) 
\end{equation} 
where $\sigma^{ij}$ is the partonic level cross section for the collisions 
$ij \rightarrow {\tilde q} {\tilde {\bar q}} ({\tilde g}{\tilde g})$ etc., where the indices 
$i,j$ run over q, $\bar q$ and g. The partonic level center of mass energy is related to the
hadronic level center of mass energy by: $\hat s~=~x_1x_2 s$.
The partonic level squark and gluino production cross 
sections $\sigma^{ij}(\hat s) $ can be obtained from the matrix element squares above
and are given by: 

\begin{small}
\begin{eqnarray*}
  \sigma^B(q_i\qb_j\to\sq\sqb) & = &
  \delta_{ij} \,\frac{n_f\pi\as^2}{\hat s}\,\bs\left[\frac{4}{27}
  -\frac{16\ms^2}{27\hat s} \right] \\
  &&{} +\delta_{ij}\,\frac{\pi\as\hat{\alpha}_s}{\hat s}\left[\bs\left(\frac{4}{27}
  +\frac{8\md^2}{27\hat s}\right) +\left(\frac{8\mg^2}{27\hat s}
  +\frac{8\md^4}{27\hat s^2} \right) 
   \Lg{\hat s+2\md^2-\hat s\bs}{\hat s+2\md^2+\hat s\bs} \right] \nonumber \\ 
  &&{}  +\frac{\pi\hat{\alpha}_s^2}{\hat s}\left[\bs\left(
  -\frac{4}{9}-\frac{4\md^4}{9(\mg^2\hat s+\md^4)}\right)
  +\left(-\frac{4}{9} -\frac{8\md^2}{9\hat s}\right) 
   \Lg{\hat s+2\md^2-\hat s\bs}{\hat s+2\md^2+\hat s\bs} \right]\nonumber \\[0.2cm]
  \sigma^B(gg\to\sq\sqb) & = &
  \frac{n_f\pi\as^2}{\hat s}\left[
  \bs \left(\frac{5}{24} +\frac{31\ms^2}{12\hat s} \right)
  +\left(\frac{4\ms^2}{3\hat s}+ \frac{\ms^4}{3\hat s^2}\right)
  \log\left(\frac{1-\bs}{1+\bs}\right) \right]\\[0.2cm]
  \sigma^B(q_i q_j\to\sq\sq) & = &
  \frac{\pi\hat{\alpha}_s^2}{\hat s}\left[\bs\left(
  -\frac{4}{9}-\frac{4\md^4}{9(\mg^2\hat s+\md^4)}\right)
  +\left(-\frac{4}{9} -\frac{8\md^2}{9\hat s}\right) 
   \Lg{\hat s+2\md^2-\hat s\bs}{\hat s+2\md^2+\hat s\bs} \right] \\ 
  & &{} + \delta_{ij}\,\frac{\pi\hat{\alpha}_s^2}{\hat s}
  \left[ \frac{8\mg^2}{27(\hat s+2\md^2)} 
   \Lg{\hat s+2\md^2-\hat s\bs}{\hat s+2\md^2+\hat s\bs} \right] \nonumber\\[0.2cm]
  \sigma^B(q\qb\to\gl\gl) & = &
  \frac{\pi\as^2}{\hat s}\, \bg\left(\frac{8}{9} +\frac{16\mg^2}{9\hat s}\right)
  \\
  & &{} +\frac{\pi\as\hat{\alpha}_s}{\hat s}\left[
  \bg\left(-\frac{4}{3}-\frac{8\md^2}{3\hat s} \right)
  +\left(\frac{8\mg^2}{3\hat s} +\frac{8\md^4}{3\hat s^2}\right)
  \Lg{\hat s-2\md^2-\hat s\bg}{\hat s-2\md^2+\hat s\bg} \right] \nonumber\\
  & &{} +\frac{\pi\hat{\alpha}_s^2}{\hat s}\left[
  \bg\left(\frac{32}{27}+\frac{32\md^4}{27(\ms^2\hat s+\md^4)}\right)
  +\left(-\frac{64\md^2}{27\hat s} -\frac{8\mg^2}{27(\hat s-2\md^2)}\right)
   \Lg{\hat s-2\md^2-\hat s\bg}{\hat s-2\md^2+\hat s\bg} \right] \nonumber\\[0.2cm]
  \sigma^B(gg\to\gl\gl) & = &
  \frac{\pi\as^2}{\hat s}\left[
  \bg\left(-3-\frac{51\mg^2}{4\hat s}\right)
  + \left(-\frac{9}{4}
  -\frac{9\mg^2}{\hat s} +\frac{9\mg^4}{\hat s^2} \right)
  \log\left(\frac{1-\bg}{1+\bg} \right) \right] \\[0.2cm]
  \sigma^B(qg\to\sq\gl) & = &
  \frac{\pi\as\hat{\alpha}_s}{\hat s}\,\left[
  \frac{\kappa}{\hat s}\left(-\frac{7}{9} -\frac{32\md^2}{9\hat s}\right)
  + \left(-\frac{8\md^2}{9\hat s}+\frac{2\ms^2\md^2}{\hat s^2}
  +\frac{8\md^4}{9\hat s^2} \right) 
  \Lg{\hat s-\md^2-\kappa}{\hat s-\md^2+\kappa} \qquad \right. \\
  & &\hphantom{\frac{\pi\as\hat{\alpha}_s}{\hat s}a} \left. {}
  + \left(-1-\frac{2\md^2}{\hat s}+\frac{2\ms^2\md^2}{\hat s^2}
  \right)  \Lg{\hat s+\md^2-\kappa}{\hat s+\md^2+\kappa} \right], \nonumber 
\end{eqnarray*}
with
\begin{eqnarray}
  \bs  = \sqrt{1-\frac{4\ms^2}{\hat s}} \qquad 
  \bg  = \sqrt{1-\frac{4\mg^2}{\hat s}} \\
  \md^2  = \mg^2 -\ms^2 \qquad
  \kappa  = \sqrt{(\hat s-\mg^2-\ms^2)^2-4\mg^2\ms^2} \\
  \as  = g_s^2/4\pi \qquad
  \hat{\alpha}_s  = \ghat^2/4\pi. 
\end{eqnarray}
\end{small}

In this paper we set the strong coupling equal to the QCD-SUSY coupling: 
$g_s = \hat{g_s}$. In our calculation we use the CTEQ6M PDF inside the proton \cite{cteq}. 
The factorization and renormalization scales are taken to be $Q=m_{\tilde{q}}, m_{\tilde{g}}$, 
(the squark and gluino masses respectively). We multiply a K factor of 1.5 to take into
account the higher order corrections.

\section{squark and gluino production From a TeV Scale Blackhole at LHC} 

If a black hole is formed at LHC, it will quickly evaporate by emitting thermal 
Hawking radiation. The emission rate in time 
for a SUSY particle with momentum $p =|\vec p|$ and 
energy $Q= \sqrt{p^2+M^2}$ can be written \cite{gram} as 
\be 
\frac{dN}{dt }= \frac{c_s \sigma_s}{8\pi^2}\frac{dp p^2}{(e^{Q/T_{BH}} \pm 1)} 
\label{thermal}
\ee 
where $\sigma_s$ is the grey body factor and $T_{BH}$ is the blackhole temperature, 
which depends on the number of extra dimensions and the TeV scale Planck mass. 
$c_s$ is the multiplicity factor.
The $\pm$ sign refers to squark and gluino respectively. The temperature of the 
blackhole is given by \cite{pp}: 
\be T_{BH}=\frac{d+1}{4\pi R_{S}} ~=~
\frac{d+1}{4 \sqrt{\pi}}~
M_P [\frac{M_P}{M_{BH}}\frac{d+2}{8\Gamma(\frac{d+3}{2})}]^{\frac{1}{1+d}} 
\ee 
where $R_{S}$ is the Schwarzschild radius of the black hole, 
$M_P$ is the TeV scale Planck mass and $d$ is the number of extra dimensions. 
The grey body factor in the geometrical approximation is given by \cite{gray4,bv,grayd}
\be 
\sigma_s = \Gamma_s 4\pi (\frac{d+3}{2})^{2/(d+1)} \frac{d+3}{d+1} R^2_{S} 
\ee 
where we take $\Gamma_s$ = 2/3 and 1/4 for spin $s$= 1/2 and 1 particles respectively. 
The SUSY energy and mass are related by $Q^2=p^2+M^2$, where $M$ is the mass of the 
SUSY particle.  From the above equation we get: 
\be 
\frac{dN}{dt dp}= \frac{c_s \sigma_s}{8\pi^2}\frac{p^2 }{(e^{\sqrt{p^2+M^2}/T_{BH}} \pm 1)}. 
\ee 
The total number of sparticles emitted by a blackhole is thus given by: 
\be N_{SUSY}= \int_0^{t_f} dt \int_0^{M_{BH}} dp 
\frac{c_s \sigma_s}{8\pi^2}\frac{p^2 }{(e^{\sqrt{p^2+M^2}/T_{BH}} \pm 1)} 
\label{NN}
\ee 
where $t_f$ is the total time taken by the blackhole to completely evaporate, 
a time which takes the form \cite{pp1}:
\be
t_f~=~\frac{C}{M_P}(\frac{M_{BH}}{M_P})^{\frac{d+3}{d+1}}
\ee
where $C$ depends on the extra dimension, polarization degrees 
of freedom etc. However, the complete determination of $t_f$
depends on the energy density present out side the blackhole
which we computed in a previous paper \cite{cooper} where we considered
the absorption of the quark-gluon plasma \cite{plasma} by TeV scale black hole at LHC.
(this time is typically about 
$10^{-27}$ sec). 
The value we use throughout this paper is $t_f$= $10^{-3}$ fm
which is normally the inverse of the TeV scale energy. 

This is the expression (Eq. {\ref{NN}) for sparticle
emission from a single blackhole of 
temperature $T_{BH}$. To obtain the SUSY production cross section from all blackholes produced in pp 
collisions at LHC we need to multiply the blackhole 
production cross section with the number of sparticles produced from a single blackhole. 
In the following we will calculate the blackhole 
production cross section in a pp collision at LHC.  The black hole production cross section 
$\sigma_{BH} $ at high energy hadronic collisions at zero 
impact parameter is given by \cite{pp,cham}: 
\bea \sigma_{BH}^{AB \rightarrow BH +X}(M_{BH}) 
= {\sum}_{ab}~ 
\int_{\tau}^1 dx_a \int_{\tau/x_a}^1 dx_b f_{a/A}(x_a, Q^2) \nonumber \\ 
f_{b/B}(x_b, Q^2) \hat{\sigma}^{ab \rightarrow BH }(\hat s) ~\delta(x_a x_b -M_{BH}^2/s). 
\label{bkt} 
\eea 
In the above expression $x_a (x_b)$ is the longitudinal momentum fraction of the parton inside 
the hadron A(B) and $\tau=\frac{M^2}{s}$, where $\sqrt s$ is the NN center of mass energy. 
Energy-momentum conservation implies $\hat s =x_ax_b s=M^2$, where $M$ is the mass of the black 
hole or string ball. We use $Q=M$ as the scale at which the 
parton distribution function is measured. ${\sum}_{ab}$ represents 
the sum over all partonic 
combinations. The black hole production cross sections in a binary partonic 
collision are given by \cite{pp} 
\bea \hat{\sigma}^{ab \rightarrow BH }(\hat s) = \frac{1}{M^2_P} [\frac{M_{BH}}{M_P}
(\frac{8\Gamma(\frac{d+3}{2})}{d+2})]^{2/(d+1)}Ê 
\label{bk3} 
\eea 
where $M_P$ and $M_{BH}$ are the Planck mass scale and black hole mass respectively. 
Again, d denotes the number of extra spatial dimensions. We use the CTEQ6M PDF to compute the 
blackhole cross section in pp collisions at LHC. The total cross section for SUSY 
production at LHC is then given by: 
\be \sigma_{SUSY} = N_{SUSY} \sigma_{BH}. 
\label{susybk}
\ee 
We now compare this total cross section for SUSY 
production via black hole resonances with the total number of squarks and gluinos which 
would be produced via pQCD processes, as explained in section II. To compare
the differential cross section we decompose the phase-space integration
in eq. (\ref{thermal}) as $d^3\vec p ~=~d^2p_t ~dp_z~=~d^2p_t~m_t~\cosh \,y~dy$
where $m_t~=\sqrt{p_t^2+m^2}$ and $p^{\mu}~=~(m_t \cosh \,y, p_x,p_y,m_t \sinh \,y)$.

\section{Results and Discussions} 

In this section we present the results of our calculation. 
First of all we present (see Fig. 2) the total blackhole cross section at LHC at $\sqrt s$
= 14 TeV pp collisions by using eq. (\ref{bkt}) with CTEQ6M \cite{cteq}
parton distribution function with factorization scale $Q=M_{BH}$. As 
black hole mass has to be higher than the Planck mass we have plotted
the blackhole cross section for blackhole mass values higher than the 
Planck mass values. The solid line is for $M_P$ = 1 TeV, and the upper 
(lower) dashed lines are for $M_P$ = 3 and 5 TeV respectively. These cross
sections are multiplied with the number of SUSY particles emitted from a single 
blackhole (see eq. (\ref{susybk})) to obtain the total SUSY production cross section
at LHC. We have choosen d=4.

In the rest of the calculation we choose following values
for the number of extra dimensions, Planck mass and blackhole mass.
The number of extra dimension we choose is d=4, the Planck mass $M_P$ =
1 (2) TeV with blackhole
mass of 3 (5) TeV and the Planck mass $M_P$ =3 TeV with blackhole
mass of 5 and 7 TeV and Planck mass $M_P$ = 5 TeV with blackhole mass
equal to 7 TeV. The typical values (in pico barns)  for the cross  
section for the production of SUSY
partners  (for example squarks)
emitted from a single black hole is around 0.1 to 0.8 depending on the
blackhole mass. In Fig. 3 we contrast the result for squark
production from the pQCD-SUSY calculation with that from thermal  
blackhole emission as a
function of squark mass in pp collisions at $\sqrt s$ = 14 TeV at LHC.
The solid 
line is squark production from the pQCD-SUSY calculation with 
$m_{\tilde{q}}/m_{\tilde{g}}$=0.8.
The upper dashed line, lower dashed line and dotted line 
are squark production from a blackhole at LHC with a number 
of extra dimensions d=4, Planck mass $M_P$= 1 TeV with blackhole mass equal to
3 TeV, $M_P$= 2 TeV with blackhole mass equal to 5 TeV and $M_P$= 3 TeV with
black hole mass equal to 5 TeV respectively. 
The upper (lower) dot-dashed 
lines are squark production from a blackhole 
at LHC with a number of extra dimensions d=4, Planck mass $M_P$= 3 (5) TeV and blackhole mass 
equal to 7 TeV respectively. It can be seen that the squark 
production from a TeV scale blackhole with a temperature of about 1 TeV does not depend 
too much on the squark mass. The pQCD-SUSY result for the production of squarks decreases 
rapidly as the mass of the squark is increased. On the otherhand, squark production from 
a black hole is not very sensitive to the squark mass as long as the blackhole temperature is very high. 
The upper dashed line is for $M_P$= 1 TeV where the blackhole temperature
is lower (as the Planck mass is lower), hence the cross section slightly 
decreases as the squark mass is increased. All other lines correspond
to high temperature (as the Planck mass is increased) and hence the
SUSY production cross section is not very sensitive to squark mass.
The production cross section of squarks from a blackhole is higher or lower than the pQCD-SUSY 
production rate for high mass squarks depending on the Planck mass and blackhole
mass values. For example if the Planck mass is around
1-3 TeV and black hole mass is around 5 TeV than SUSY production from
blackhole is larger for squarks of mass greater than about 300 GeV.

In Fig. 4 we contrast the result for gluino 
production from the pQCD-SUSY calculation with that from thermal blackhole emission as a function of gluino mass in pp collisions at $\sqrt s$ = 14 TeV at LHC. 
The solid line is gluino production from the pQCD-SUSY calculation with 
$m_{\tilde{q}}/m_{\tilde{g}}$=0.8.
The upper dashed line, lower dashed line and dotted line 
are gluino production from a blackhole at LHC with a number 
of extra dimensions d=4, Planck mass $M_P$= 1 TeV with blackhole mass equal to
3 TeV, $M_P$= 2 TeV with blackhole mass equal to 5 TeV and $M_P$= 3 TeV with
black hole mass equal to 5 TeV respectively. 
The upper (lower) dot-dashed lines are gluino production from a blackhole 
at LHC with a number of extra dimensions d=4, Planck mass $M_P$= 3 (5) TeV 
and blackhole mass 
equal to 7 TeV respectively. The pQCD-SUSY result for the
production of gluinos decreases rapidly as the mass of the gluino is increased. On the otherhand, gluino 
production from a black hole is not very sensitive to the gluino mass as long 
as the blackhole temperature is very high. 
The upper dashed line is for $M_P$= 1 TeV where the blackhole temperature
is lower (as the Planck mass is lower), hence the cross section slightly 
decreases as the gluino mass is increased. All other lines correspond
to high temperature (as the Planck mass is increased) and hence the
SUSY production cross section is not very sensitive to gluino mass.
The production cross section of gluinos from a blackhole is higher or 
lower than the pQCD-SUSY 
production rate for high mass gluinos depending on the Planck mass and blackhole
mass values. For example if the Planck mass is around
1-3 TeV and black hole mass is around 5 TeV than SUSY production from
blackhole is larger for gluinos of mass greater than about 400 GeV.

In Fig.5 we present the rapidity distribution results for squark 
production both from pQCD-SUSY processes and from blackholes.  The solid line is the squark 
production cross section from direct pQCD-SUSY 
production processes as a function of rapidity at LHC at $\sqrt s$ = 14 TeV pp collisions.
Here we have taken $m_{\tilde{q}}/m_{\tilde{g}} = 0.8$. 
The rapidity range covered is from -3 to 3. 
The upper (lower) dashed line 
is the rapidity distribution of squark production from a blackhole 
at LHC with a number of extra dimensions d=4, Planck mass $M_P$= 2 (3) 
TeV and blackhole mass equal to 5 (7) TeV respectively. 
The production cross section from a blackhole is higher or lower than the pQCD-SUSY 
production rate for high mass squarks depending on the Planck mass and blackhole
mass values. For example if the Planck mass is around
1-3 TeV and black hole mass is around 5 TeV than SUSY production from
blackhole is larger for squarks of mass equal to 500 GeV.

Similarly in Fig.6 we present the rapidity distribution results for gluino 
production, both from pQCD-SUSY processes and from blackholes. The solid line is the gluino 
production cross section from direct pQCD-SUSY 
production processes as a function of rapidity at LHC at $\sqrt s$ = 14 TeV pp collisions. 
Here we have taken the gluino mass to be 500 GeV with $m_{\tilde{q}}/m_{\tilde{g}} = 0.8$. 
The rapidity range covered is from -3 to 3. 
The upper (lower) dashed line 
is the rapidity distribution of gluino production from a blackhole 
at LHC with a number of extra dimensions d=4, Planck mass $M_P$= 2 (3) 
TeV and blackhole mass equal to 5 (7) TeV respectively. 
The production cross section from a blackhole is higher or lower than the pQCD-SUSY 
production rate for high mass gluino depending on the Planck mass and blackhole
mass values. For example if the Planck mass is around
1-3 TeV and black hole mass is around 5 TeV than SUSY production from
blackhole is larger for gluino of mass equal to 500 GeV.

In Fig.7 we present $d\sigma/dp_t$ for the squark production cross section, 
both from pQCD-SUSY processes and from blackholes. The solid line is the squark
production cross section from direct pQCD-SUSY 
production processes as a function of $p_t$ at LHC at $\sqrt s$ = 14 TeV pp collisions. 
Here we have taken the mass to be 500 GeV with $m_{\tilde{q}}/m_{\tilde{g}} = 0.8$. 
The upper dashed line is the $p_t$ distribution of squark production from a blackhole 
at LHC with a number of extra dimensions d=4, Planck mass $M_P$= 2 TeV, 
blackhole mass equal to 5 TeV and squark mass equal to 500 GeV. 
The lower dashed line is the $p_t$ distribution of squark production from a blackhole 
at LHC with a number of extra dimensions d=4, Planck mass $M_P$= 3 TeV, 
blackhole mass equal to 7 TeV and squark mass equal to 500 GeV. 
The dotted and dot-dashed lines are the similar curves but for squark mass equal to 1 TeV.
It can be seen clearly from the figure that $d\sigma/dp_t$ 
for squark production from a blackhole increases
as $p_t$ is increased (in all lines except solid line) 
up to $p_t$ around 1 TeV (for squark mass equal to 500 GeV)
or more (for squark mass equal to 1 TeV) which is in sharp contrast to the
pQCD predictions where the $d\sigma/dp_t$ decreases as $p_t$ is increased
up to 1 TeV. Similar results are obtained for gluino production, and this is
shown in Fig.8. This increase in $d\sigma/dp_t$ 
is not only true for SUSY production but also true for any particles emitted
from the blackhole as long as the temperature of the blackhole is large ($\sim$ TeV).
This is because it is coming from the phase-space ($d^3p$ integration
as at high temperature the decrease of thermal distribution w.r. to $p_t$
is not very rapid.
Hence observation of an increase in $d\sigma/dp_t$ (as $p_t$ is increased
up to high $p_t$ value $\sim$ 1 TeV or more)
for any particles might be a clear signature of black hole production at LHC. 

To conclude, if the fundamental Planck scale is near a TeV, then we should
expect to see TeV scale black holes at the LHC.
Similarly, if the scale of supersymmetry breaking is sufficiently 
low, then we might expect to see light supersymmetric particles in the next 
generation of colliders. If the mass of the supersymmetric particle is of order  
a TeV and is comparable to the temperature of a typical TeV scale 
black hole, then such sparticles will be copiously 
produced via Hawking radiation: 
The black hole will act as a resonance for 
sparticles, among other things. In this paper we compared various signatures for SUSY 
production at LHC, and we contrasted the situation where the sparticles are produced 
directly via parton fusion processes
with the situation where they are produced indirectly through black hole resonances. 
We found that black hole resonances provide a larger source for heavy mass SUSY 
(squark and gluino) production than the direct pQCD-SUSY production via parton 
fusion processes depending on the values of the Planck mass and blackhole mass.
Hence black hole production at LHC may indirectly act as a dominant channel for SUSY production. 
We also found that the differential cross section $d\sigma/dp_t$ for 
SUSY production increases as a function of the $p_t$ (up to $p_t$ equal to about 1
 TeV or higher) of the SUSY particles (squarks and
gluinos), which is in sharp contrast with the pQCD predictions where the differential cross section
$d\sigma/dp_t$ decreases as $p_t$ increases for high $p_t$ about 1 TeV
or more. This is a feature for any 
particle emission from TeV scale blackhole as long as the
temperature of the blackhole is very high ($\sim $ TeV). 
Hence measurement of increase of $d\sigma/dp_t$ with 
$p_t$ for $p_t$ up to about 1 TeV or more
for final state particles might be a useful 
signature for blackhole production at LHC. 

\acknowledgements 
We thank Jack Smith, George Sterman, and John Terning
for useful discussions. This work was supported in part by the National Science
Foundation grant PHY-0098527 and Depertment of Energy, under contract W-7405-ENG-36.

\begin{figure}
\begin{center} 
\vspace*{-2.3cm} 
\hspace*{-0cm} 
\epsfig{file=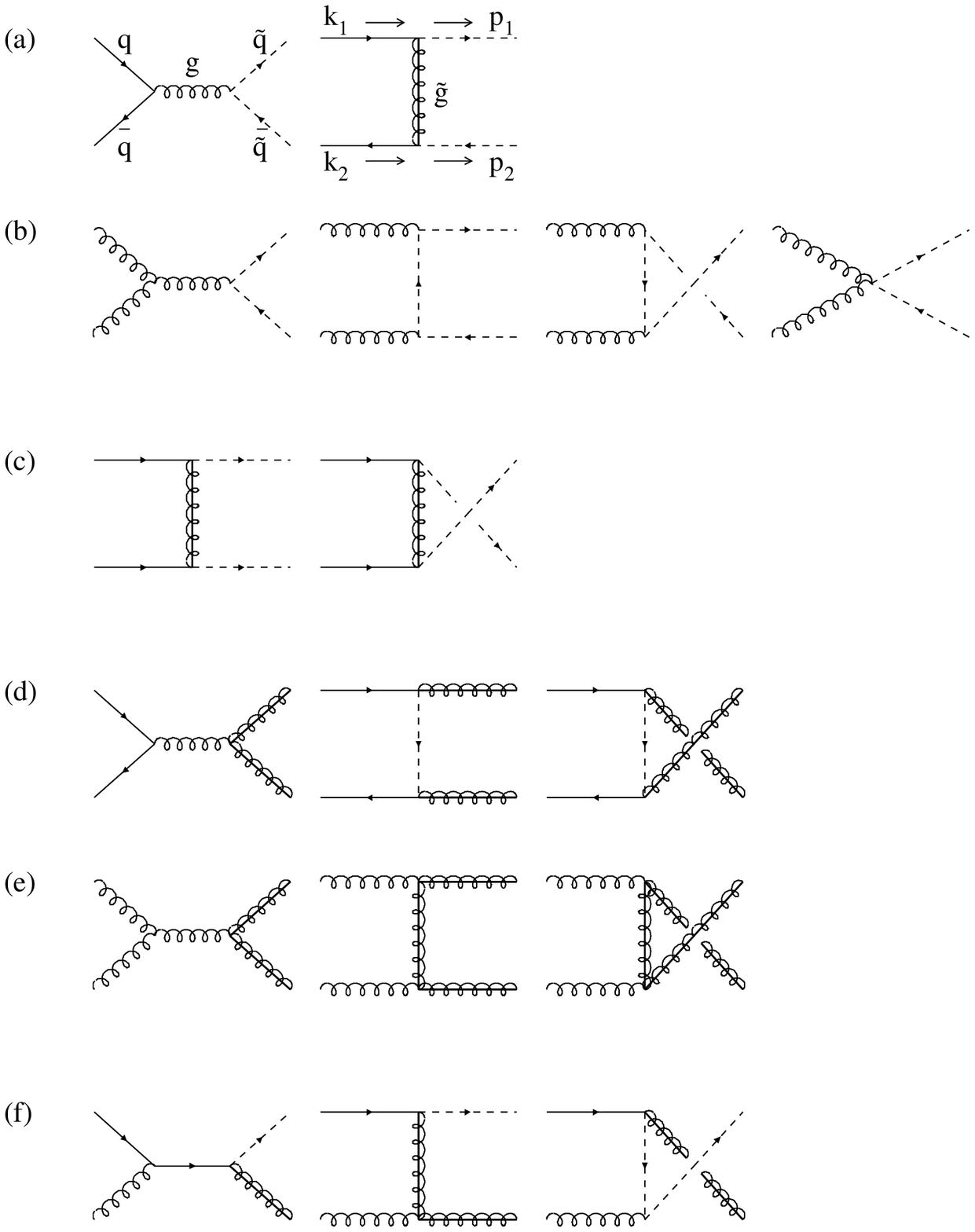,width=17cm} 
\vspace*{-2.5cm} 
\end{center} 
\caption{Feynman diagrams for the production of squarks and 
gluinos in lowest order. } 
\label{fig1} 
\end{figure} 

\begin{figure} 
\includegraphics[width=3in]{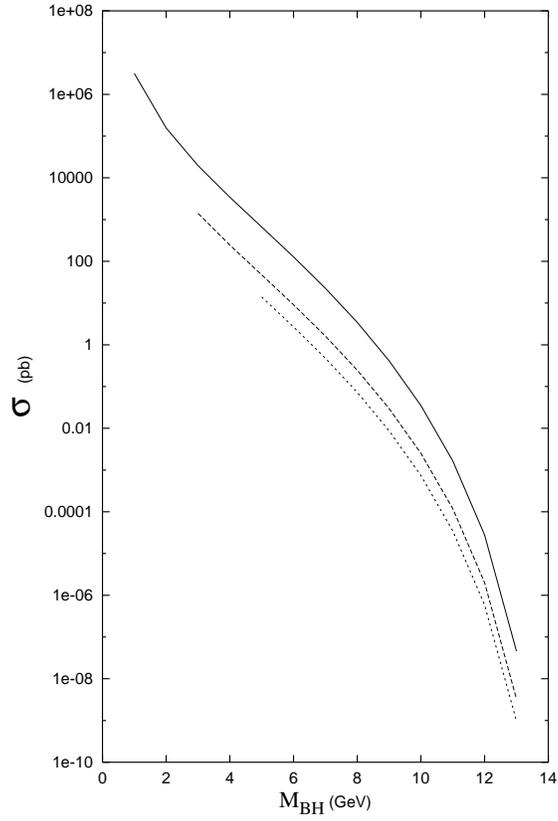} 
\caption{ Total cross section for blackhole production at LHC.
The solid line is for $M_P$=1 TeV and the upper (lower) dashed line is
for $M_P$= 3 (5) TeV. } 
\label{fig2} 
\end{figure} 

\begin{figure} 
\includegraphics[width=4in]{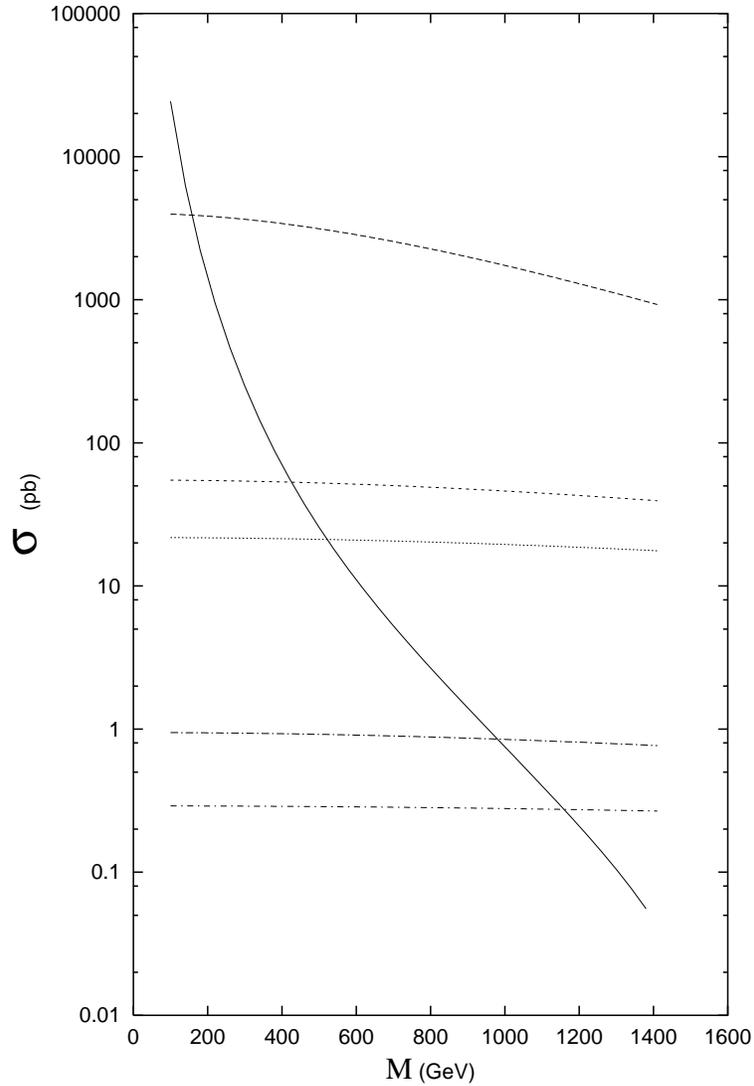} 
\caption{ Total cross section for squark production at LHC as a 
function of squark mass.
The solid line is from direct pQCD processes and other lines 
are from blackhole decay depending on the Planck mass and blackhole mass
as explained in the text. The top line is for blackhole mass equal to 3 TeV,
next two almost horizontal lines are for a blackhole mass of 
5 TeV and the bottom two are for a blackhole mass of 7 TeV.} 
\label{fig3} 
\end{figure} 

\begin{figure} 
\includegraphics[width=3in]{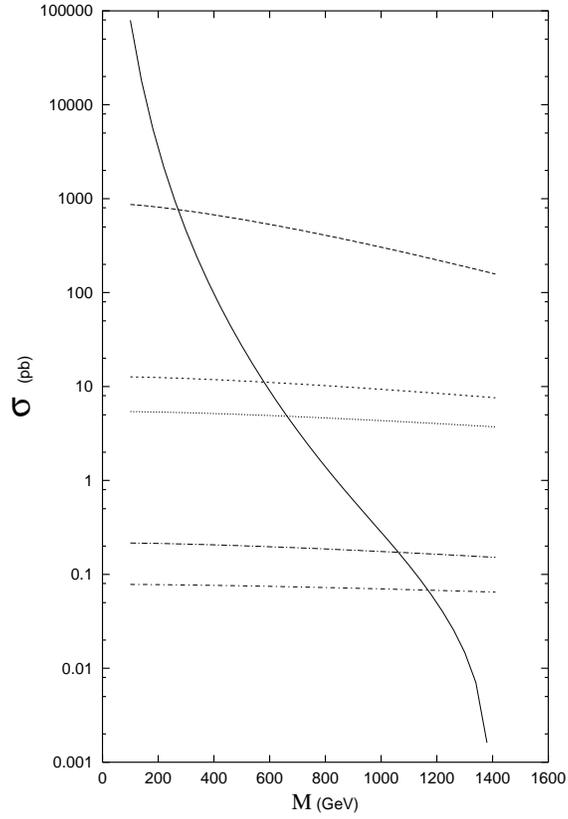} 
\caption{ Total cross section for gluino production at LHC as a function
of gluino mass. The solid line is from direct pQCD processes and other lines 
are from blackhole decay depending on the Planck mass and blackhole mass
as explained in the text. Values of the blackhole mass are as in  Fig. \ref{fig3} .} 
\label{fig4} 
\end{figure} 

\begin{figure} 
\includegraphics[width=3in]{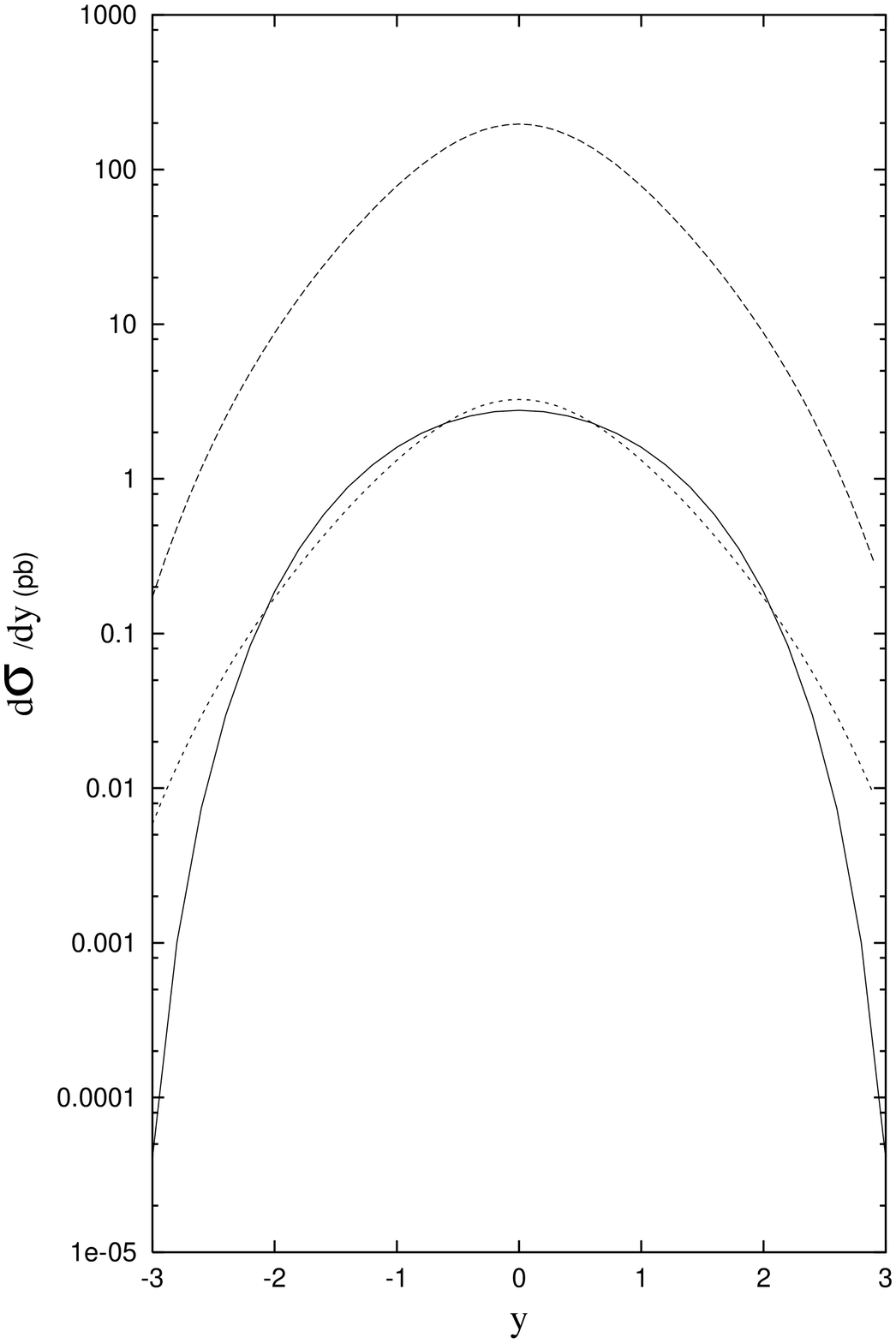} 
\caption{ The rapidity distribution of squark production cross section at LHC.
The solid line is from direct pQCD processes and other lines 
are from blackhole decay depending on the Planck mass and blackhole mass
as explained in the text. The squark mass is choosen to be equal to 500 GeV.
The upper (lower) dashed line corresponds to a blackhole mass of 5 (7) TeV.} 
\label{fig5} 
\end{figure} 

\begin{figure} 
\includegraphics[width=3in]{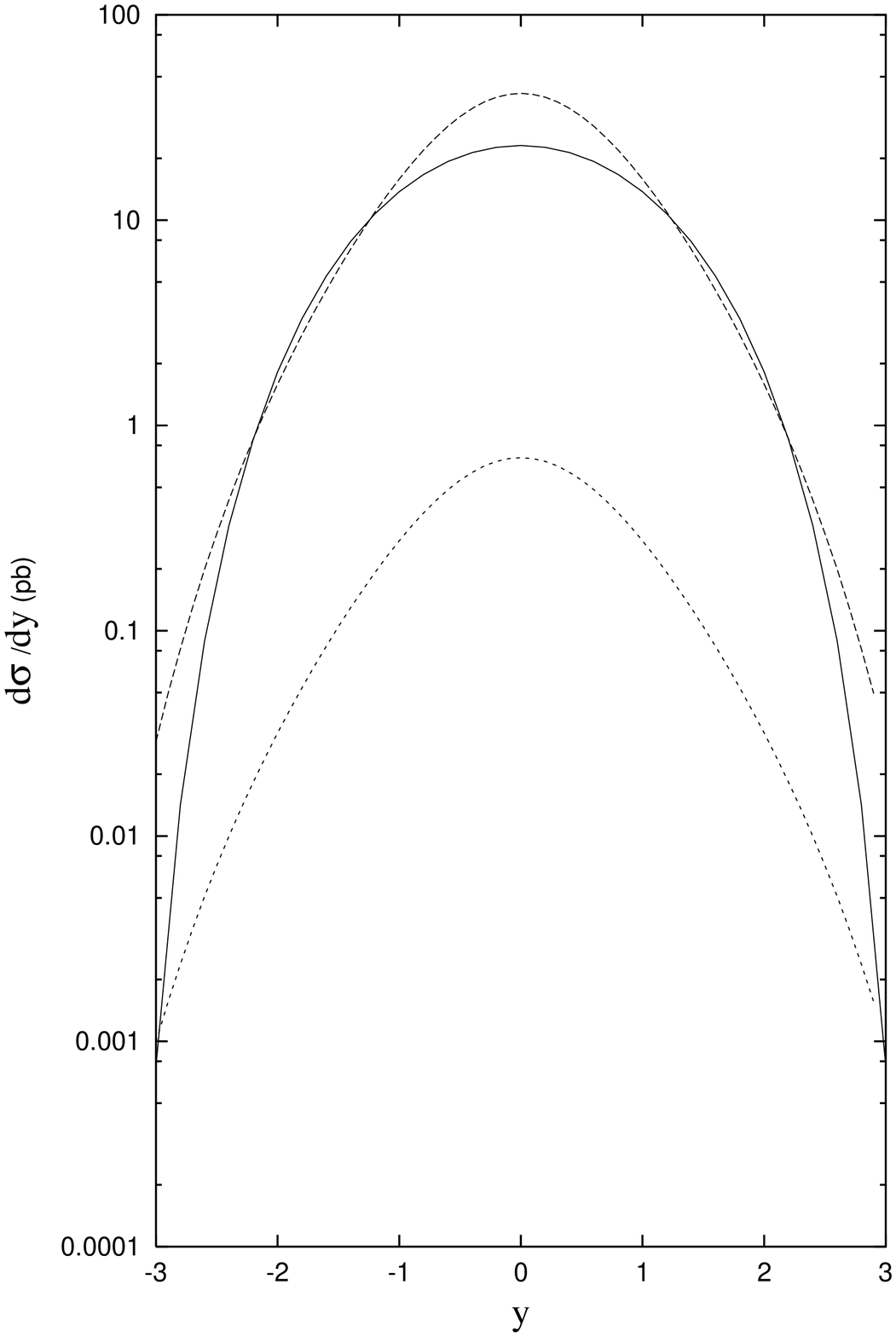} 
\caption{ The rapidity distribution of gluino production cross section at LHC.
The solid line is from direct pQCD processes and other lines 
are from blackhole decay depending on the Planck mass and blackhole mass
as explained in the text. The gluino masses are choosen to be equal to 500 GeV.
The upper (lower) dashed line corresponds to a blackhole mass of 5 
(7) TeV.} 
\label{fig6} 
\end{figure} 

\begin{figure} 
\includegraphics[width=3in]{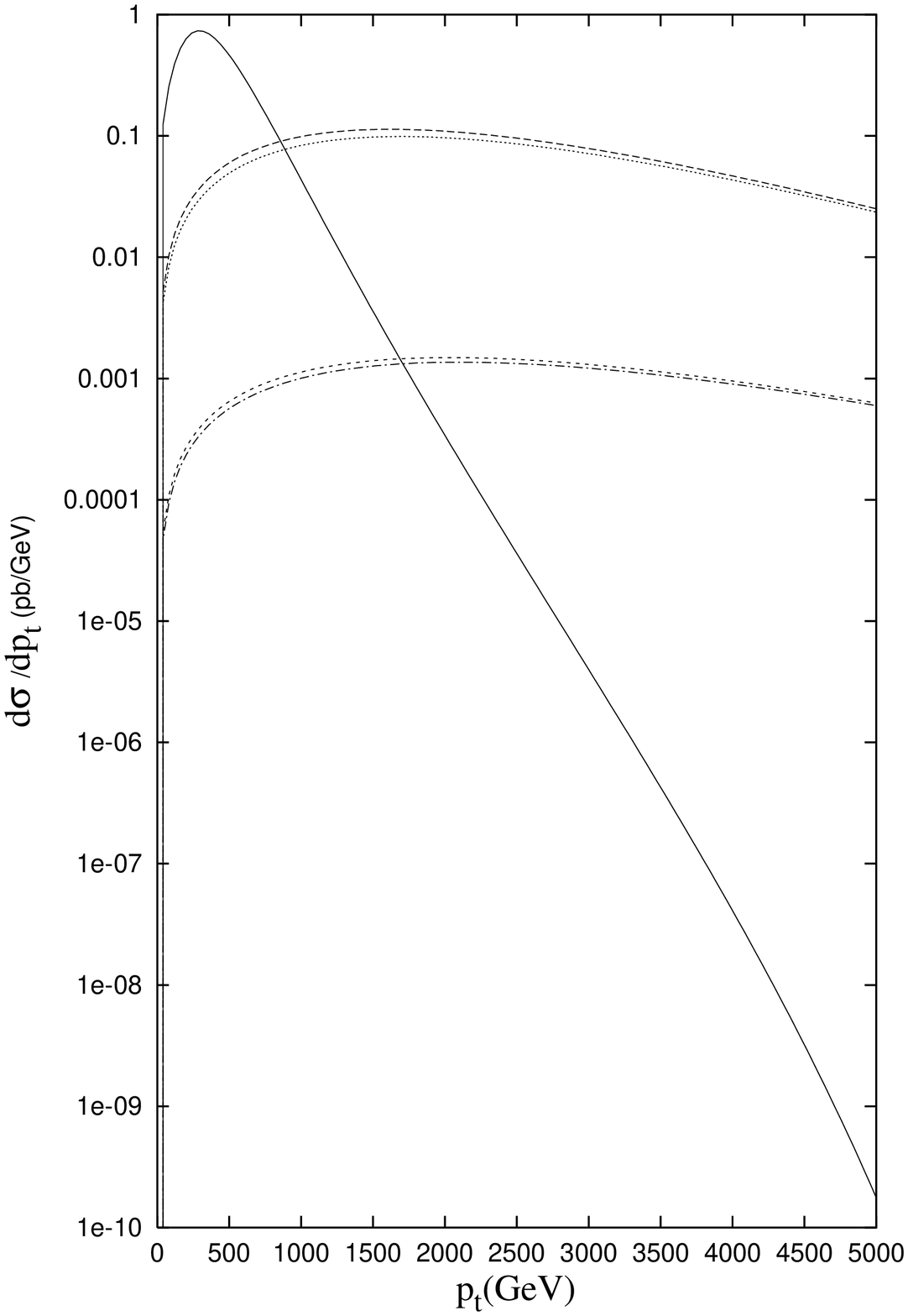} 
\caption{ The $p_t$ distribution of squark production cross section at LHC.
The solid line is from direct pQCD processes and other lines 
are from blackhole decay depending on the Planck mass and blackhole mass
as explained in the text. The squark masses are choosen to be equal to 500 GeV
and 1 TeV. The upper (lower) other lines correspond to a blackhole mass of 
5 (7) TeV.} 
\label{fig7} 
\end{figure} 

\begin{figure} 
\includegraphics[width=3in]{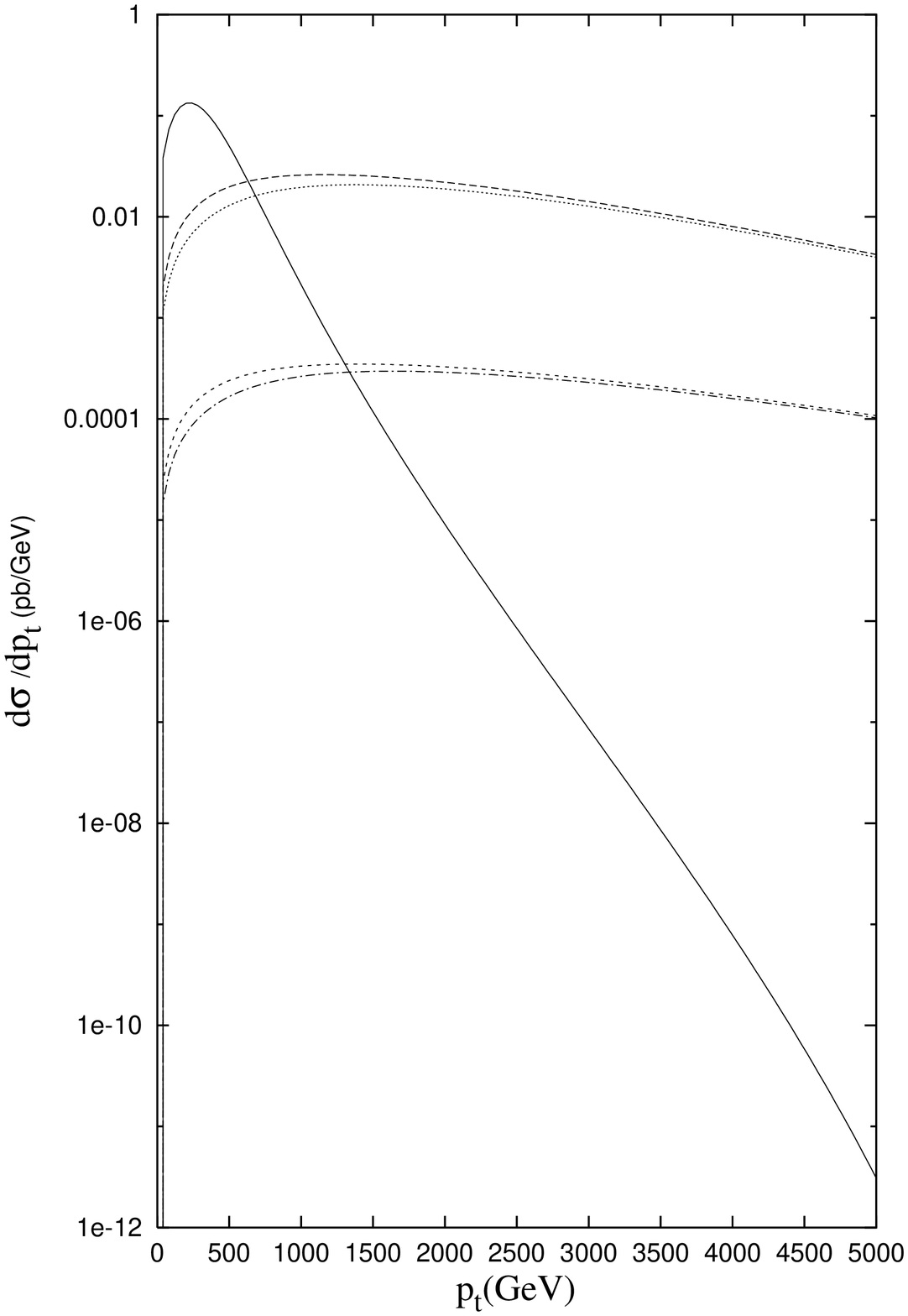} 
\caption{ The $p_t$ distribution of gluino production cross section at LHC.
The solid line is from direct pQCD processes and other lines 
are from blackhole decay depending on the Planck mass and blackhole mass
as explained in the text. The gluino mass is choosen to be equal to 500 GeV.
The upper (lower) other lines correspond to a blackhole mass of 5 (7) TeV.} 
\label{fig8} 
\end{figure} 


\begin{thebibliography}{aaaaaaa}

\bibitem{folks}
N. Arkani-Hamed, S. Dimopoulos and G. Dvali,
Phys. Lett. {\bf B429}, 263 (1998);
I. Antoniadis, N. Arkani-Hamed, S. Dimopoulos and G. Dvali,
Phys. Lett. {\bf B436}, 257 (1998);
L.~Randall and R.~Sundrum, Phys. Rev. Lett. {\bf 83}, 3370 (1999);
L.~Randall and R.~Sundrum, Phys. Rev. Lett. {\bf 83}, 4690 (1999).

\bibitem{gravr} G. F. Giudice, R. Rattazzi and J. D. Wells, Nucl. Phys. B
544 (1999) 3; L. Vacavant and I. Hinchliffe, 
J. Phys. G27 (2001) 1839; G. C. Nayak, hep-ph/0211395; 
Y. A. Kubyshin, hep-ph/0111027; 
S.~B.~Bae et al., Phys.\ Lett.\ B {\bf 487}, 299 (2000);
S. C. Park, H. S. Song and J. Song, Phys. Rev. D63 (2001) 077701;
K.~Cheung, Phys.\ Rev.\ D {\bf 63}, 056007 (2001).

\bibitem{gravr1} 
W.~D.~Goldberger and M.~B.~Wise,
Phys.\ Rev.\ Lett.\  {\bf 83}, 4922 (1999);
Phys.\ Lett.\ B {\bf 475}, 275 (2000);
Phys.\ Rev.\ D {\bf 60}, 107505 (1999);
C.~Cs\'aki, M.~Graesser, L.~Randall and J.~Terning,
Phys.\ Rev.\ D {\bf 62}, 045015 (2000);
C.~Cs\'aki, M.~L.~Graesser and G.~D.~Kribs,
Phys.\ Rev.\ D {\bf 63}, 065002 (2001);
Phys.\ Rev.\ D {\bf 62}, 067505 (2000);
C.~Csaki, J.~Erlich and J.~Terning,
Phys.\ Rev.\ D {\bf 66}, 064021 (2002);
C.~Csaki, J.~Erlich, T.~J.~Hollowood and Y.~Shirman,
Nucl.\ Phys.\ B {\bf 581}, 309 (2000);
E.~A.~Mirabelli, M.~Perelstein and M.~E.~Peskin,
Phys.\ Rev.\ Lett.\  {\bf 82}, 2236 (1999);
D.~Dominici, B.~Grzadkowski, J.~F.~Gunion and M.~Toharia,
arXiv:hep-ph/0206192;
T.~Han, G.~D.~Kribs and B.~McElrath,
Phys.\ Rev.\ D {\bf 64}, 076003 (2001);
M.~Chaichian, A.~Datta, K.~Huitu and Z.~h.~Yu,
Phys.\ Lett.\ B {\bf 524}, 161 (2002);
J.~L.~Hewett and T.~G.~Rizzo, 
{\tt hep-ph/0202155};
For a review see G.~D.~Kribs,
{\tt hep-ph/0110242}.

\bibitem{large} A. Chamblin, S. W. Hawking and H. S. Reall,
Phys. Rev. {\bf D61}, 065007(2000);
R. Emparan, G. T. Horowitz and R. C. Myers, JHEP {\bf 0001},
07(2000); N. Dadhich, R. Maartens, P. Papadopoulos and
V. Rezania, Phys. Lett. B487 (2000) 1; A. Chamblin, H. Reall, H. Shinkai and
T. Shiromizu, Phys. Rev. {\bf D63}, 064015 (2001); P. Kanti and K. Tamvakis,
Phys. Rev. {\bf D65}, 084010 (2002); C. Germani and R. Maartens,
Phys. Rev. {\bf D64}, 124010 (2001); I. Giannakis and H. Ren,
Phys. Rev. {\bf D63}, 125017 (2001); R. Casadio and L. Mazzacurati,
gr-qc/0205129; P. Kanti, I. Olasagasti and K. Tamvakis, Phys. Rev. D66
(2002) 104026.

\bibitem{ppbf} T. Banks and W. Fischler, hep-th/9906038.

\bibitem{pp} S.~Dimopoulos and G.~Landsberg,
Phys.\ Rev.\ Lett.\  {\bf 87}, 161602 (2001).


\bibitem{pp1} S.~B.~Giddings and S.~Thomas,
Phys.\ Rev.\ D {\bf 65}, 056010 (2002).

\bibitem{pp2} S.~B.~Giddings,
in {\it Proc. of the APS/DPF/DPB Summer Study on the Future of Particle Physics
(Snowmass 2001) } ed. R.~Davidson and C.~Quigg, hep-ph/0110127.

\bibitem{pp3} D.~M.~Eardley and S.~B.~Giddings, Phys. Rev. D66 (2002) 044011.


\bibitem{ag}  L. Anchordoqui and H. Goldberg, Phys. Rev. {\bf D65} 047502,
2002.

\bibitem{ppch}  R. Casadio and B. Harms, Int. J. Mod. Phys. A17 (2002) 4635.

\bibitem{ppk}  K. Cheung, Phys. Rev. D66 (2002) 036007; K. Cheung, Phys.
Rev. Lett. 88 (2002) 221602; 
K. Cheung and Chung-Hsien Chou, Phys. Rev. D66 (2002) 036008.

\bibitem{ppu} Y. Uehara, Mod. Phys. Lett. A17 (2002) 1551.

\bibitem{park} Seong Chan Park and H.S.Song, hep-ph/0111069.

\bibitem{hof} M. Bleicher, S. Hofmann, S. Hossenfelder, H. Stoecker, 
Phys. Lett. B548 (2002) 73; S. Hossenfelder, S. Hofmann, M. Bleicher, 
H. Stoecker, Phys. Rev. D66 (2002) 101502. 

\bibitem{more} I. Mocioiu, Y. Nara and I. Sarcevic, hep-ph/0301073;
V. Frolov and D. Stojkovic, gr-qc/0301016; gr-qc/0211055; Phys. Rev. D66
(2002) 084002; D. Ida and S. C. Park, hep-th/0212108; B. Kol, hep-ph/0207037;
T.~Han, G.~D.~Kribs and B.~McElrath, hep-ph/0207003;

\bibitem{cham} A. Chamblin and G. C. Nayak, Phys. Rev. D66 (2002) 091901.

\bibitem{cooper} A. Chamblin, F. Cooper and G. C. Nayak, Phys. Rev. D69 (2004) 065010.

\bibitem{pp4} L. A. Anchordoqui, J. L. Feng, H. Goldberg and A. D. Shapere,
Phys. Rev. D66 (2002) 103002; J.~L.~Feng and A.~D.~Shapere, 
Phys. Rev. Lett. 88 (2002) 021303; 
L. A. Anchordoqui, T. Paul, S. Reucroft and J. Swain,
hep-ph/0206072.

\bibitem{pp5} R.~Emparan, M.~Masip and R.~Rattazzi, Phys. Rev. D65 (2002)
064023.

\bibitem{pp6} A.~Ringwald and H.~Tu,
Phys.\ Lett.\ B {\bf 525}, 135 (2002).

\bibitem{sball} 
It is worth noting that the string ball picture for the endpoint of 
Hawking radiation only really makes sense at weak string coupling.

\bibitem{lands} G. Landsberg, Phys. Rev. Lett. 88 (2002) 181801.

\bibitem{been} W. Beenakker {\it et al.}, Nucl. Phys. B492 (1997) 51.

\bibitem{cteq} J. Pumplin {\it et al.} J. High Energy Phys. 07 (2002) 012.

\bibitem{gram} T. Han, G. D. Kribs and B. McElrath, Phys. Rev. Lett. 90 (2003) 031601.

\bibitem{gray4} L. Anchordoqui and H. Goldberg, hep-ph/0209337 

\bibitem{bv} R. Emparan, G. T. Horowitz and R. C. Myers, Phys. Rev.  Lett. 85 (2000) 499.

\bibitem{grayd} P. Kanti, J. March-Russell, hep-ph/0212199. 

\bibitem{plasma} F. Cooper, E. Mottola and G. C. Nayak, hep-ph/0210391,
Phys. Lett. B 555 (2003) 181; G. C. Nayak {\it et al.}, Nucl. Phys. A687 (2001) 457;
R. S. Bhalerao and G. C. Nayak, Phys. Rev. C61 (2000) 054907; G. C. Nayak and 
V. Ravishankar, Phys. Rev. C58 (1998) 356; Phys. Rev. D55 (1997) 6877.

\end{thebibliography}
\end{document}